# Thermoplasmonics: Quantifying Plasmonic Heating in Single Nanowires

*Joseph B. Herzog*[1,2,#], *Mark W. Knight,*[3,4,#] *Douglas Natelson,*[1,3,a)]

[1]Department of Physics and Astronomy, Rice University, Houston, Texas, 77005, United States

[2]Department of Physics, University of Arkansas, Fayetteville, Arkansas, 72701, United States

[3]Department of Electrical and Computer Engineering, [4]Laboratory for Nanophotonics,

ABSTRACT: Plasmonic absorption of light can lead to significant local heating in metallic nanostructures, an effect that defines the sub-field of thermoplasmonics and has been leveraged in diverse applications from biomedical technology to optoelectronics. Quantitatively characterizing the resulting local temperature increase can be very challenging in isolated nanostructures. By measuring the optically-induced change in resistance of metal nanowires with a transverse plasmon mode, we quantitatively determine the temperature increase in single nanostructures, with the dependence on incident polarization clearly revealing the plasmonic heating mechanism. Computational modeling explains the resonant and nonresonant contributions to the optical heating and the dominant pathways for thermal transport. These

---

[a)] Author to whom correspondence should be addressed. E-mail: natelson@rice.edu

results, obtained by combining electronic and optical measurements, place a bound on the role of optical heating in prior experiments, and suggest design guidelines for engineered structures meant to leverage such effects.

Thermoplasmonics encompasses the electronic and ionic heating in plasmonic structures under illumination and is an emerging field of critical interest[1–4] due to applications in photothermal therapeutics,[5–8] drug release,[9] thermal-optical data storage,[10] enhanced catalysis,[11–13] magnetic recording,[14,15] solar thermal energy harvesting,[16–18] and optoelectronic devices.[19,20] While under continuous illumination, energy is transferred from the incident radiation into the electronic system via both direct absorption and resonant coupling into plasmon modes; this energy is eventually transferred to the lattice via the electron-phonon coupling. In the steady state, the electronic and vibrational degrees of freedom are often adequately described by quasiequilibrium distribution functions with effective electronic and ionic temperatures. Those effective temperatures are determined by the balance between the incoming flux of energy from the optical interactions and the outgoing transfer of heat via conduction, radiation, and in a fluid environment, convection. Determining those effective temperatures with precision and accuracy is usually difficult,[21–26] particularly for discrete plasmonically-active nanoparticles.[27] Thermal microscopy techniques have been demonstrated to probe temperature changes of small volumes including a thermographic phosphor technique,[28] scanning thermal microscopy,[29] and other methods.[30–32] This work combines electronic measurements with precise illumination to detect temperature changes isolated to the nanoscale volume of a single nanowire.

This study examines the relative importance of plasmon and direct absorption in the heating of metal nanostructures. By contacting plasmonically active metal nanowires with extended contact pads, similar to those in nanojunction experiments,[33–35] we are able to use the temperature dependence of the wire's electrical resistivity to determine the light-induced temperature increase. The well-defined local surface plasmon resonances (LSPRs) of the nanowires, which resonate transverse to the wire axis, lead to a dipolar dependence of the optical heating on the incident polarization. This polarization dependence is consistent with previous plasmonic studies and calculations,[35] and the magnitude of the heating as a function of wire geometry is consistent with finite element modeling of the optical interactions and the thermal transport. Under illumination conditions like those used in previous work,[33–35] the optically driven temperature increase is approximately 2 K at a substrate temperature of 80 K, which is comparable to results seen in other thermoplasmonic studies.[28] The dependence of the optically induced temperature elevation on the substrate's structure and the wire geometry confirm that the dominant path for heat flow out of the wire is via electronic thermal conduction in the metal, and then into the substrate from the large contacts via phonon thermal conduction. We discuss the implications of these results for nanojunction measurements and the prospect of plasmon-based bolometric devices.[36]

The plasmonic devices studied in this work are Au nanowires on thermally oxidized silicon substrates. Electron beam lithography was used to pattern arrays of nanowires linking two larger electrodes on substrates coated with either 0.2 μm or 2 μm of $SiO_2$. The nanowires in this work had a nominal length of 600 nm, and the widths were varied from 100 nm to 160 nm. After lithography and developing the exposed resist, metal was deposited using an electron beam evaporator: 1 nm of Ti and 13 nm of Au, followed by lift-off with acetone. A nanowire with the standard design is shown in Figure 1a and a larger field of view showing the nanowire and electrodes is shown in Figure 1b. Another design variation involved increasing the nanowire length, as shown in Figure 1c. After the nanowires with triangular electrodes are fabricated, contact pads hundreds of microns in extent that overlapped the triangular electrodes were deposited using a shadow mask and a subsequent metal evaporation. Finally, these electrodes were wire-bonded to a chip carrier for electronic characterization.

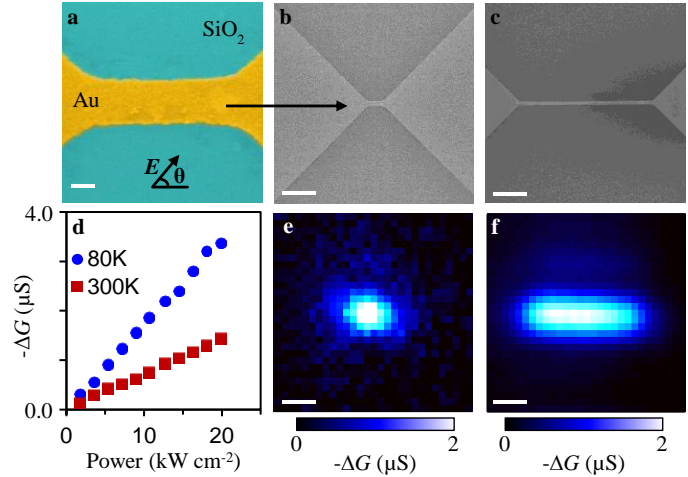

**Figure 1.** Scanning electron microscope (SEM) images (a-c) of representative devices and (d-f) typical optical conduction responses. (a) Colorized SEM image of a Au nanowire on 0.2 μm of SiO$_2$, with reference angle for polarization studies. (b) Typical overall device geometry. (c) Elongated design. (d) Change in conductance, −ΔG, due to optical heating as a function of incident laser intensity with 1.8 μm FWHM Gaussian spot centered on the nanowire, measured at room and low temperatures ($\theta = 90°$). (e-f) Room temperature diffraction-limited scans of the change in conductance due to optical heating for (e) standard and (f) elongated geometries with $\theta = 90°$. Images are generated by raster scanning the laser across the samples pictured in the SEM images (b-c). Scale-bars for each image are (a) 100 nm, (b-c) 1 μm, and (e-f) 1 μm.

Electrical measurements under focused illumination were used to characterize the devices and quantify the plasmonic heating (Figure S1). A 785 nm laser was rastered across the sample using a telescopic scanning lens system before the 50x (N.A. = 0.7) objective. The maximum laser intensity at the sample was 20 kW/cm$^2$ and could be adjusted using a variable neutral density filter. The Gaussian laser beam was focused to a minimum FWHM of 1.8 μm. Rotating a half-wave plate controlled the incident laser polarization (with $\theta$ defined in Fig. 1a), and an optical chopper modulated the incident light at a frequency of 297 Hz. A voltage bias, $V_{dc}$, as large as 0.3 V was applied across the sample using analog outputs of a lock-in amplifier. The signal from the chopper was used as the external reference for a lock-in amplifier, which measured the optically induced change in current, $\Delta I$, transduced by a current preamplifier. The change in conduction is defined by $\Delta G = \Delta I/V_{dc}$, and the overall 2-terminal conductance of each device without the presence of the laser was also measured with lock-in techniques to determine the dark conductance as a function of temperature. The lock-in additionally generated a 2.0 kHz, 10 mV RMS ac excitation that was summed with the dc bias and applied to the device. The first harmonic response at 2.0 kHz ($dI/dV$) was measured to determine the differential conductance of each device. All measurements were performed under vacuum in an optical cryostat with a built-in sample stage heater for temperature dependent measurements.

Scanning the laser across a single nanowire while measuring change in conductance, $\Delta G$, revealed a diffraction-limited "hotspot" at the center of the device (Figure 1e). A photothermally induced decrease in conductance occurred when the laser illuminated the nanowire that limits the total device conductance. With the standard nanowire geometry, the laser was positioned at the hotspot of the device and power dependent measurements of the change in conductance were performed at 300 K and 80 K. These results show that $-\Delta G$ increased linearly with increasing laser power (Figure 1d). Measurements were also performed on elongated devices with lengths of 5 μm, significantly exceeding the laser FWHM (Figure 1f), which confirmed that the $\Delta G$ hotspot is localized to the nanowire where the plasmon resonance exists and the current density is maximized. Both scans, Figure 1e-f, were taken at room temperature, using the full 20 kW/cm$^2$ laser intensity, and with transverse polarization ($\theta = 90°$). The change in conductance, $\Delta G$, results from the increase in nanowire resistance as a result of optically induced heating.

To verify the plasmonic contribution to the heating response, $-\Delta G$ as a function of incident laser polarization, $\theta$, was measured with the laser centered on the nanowire while varying substrate temperature, wire width, and oxide thickness. These results are plotted in Figure 2. The transverse (longitudinal) polarization, $E_T$ ($E_L$), corresponds to and $\theta = 90°$ ($0°$), respectively. These measurements were performed with substrate temperatures of 300 K and 80 K on a 130 nm wide nanowire on 2 μm oxide. The polarization dependences at the two substrate temperatures are proportional, with 80K having a $-\Delta G$ that is twice as large as the $-\Delta G$ at 300K, as shown in Figure 2a.

The polarization dependence of a device on 0.2 μm oxide was also measured at 80 K (Figure 2b). The dependence of the photoresponse on the incident polarization angle is proportional to that observed on the thicker oxide, with both having a polarization ratio $\frac{\Delta G_{E_T}}{\Delta G_{E_L}} \sim 2$. However, the magnitude of $-\Delta G$ for the device with the thinner oxide is 25% smaller. Polarization measurements of 53 standard geometry Au devices on the 0.2 μm oxide thickness substrate show polarization ratios, $\frac{\Delta G_{E_T}}{\Delta G_{E_L}}$, ranging from 1.75 – 2.5. The distribution of ratios within this range is the result of variations in the geometry, primarily the nanowire thickness.

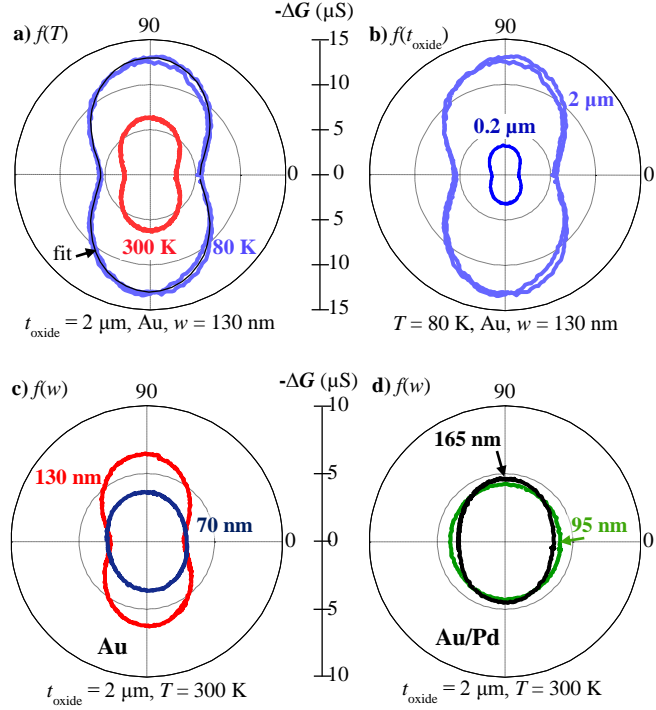

**Figure 2.** Change in conduction due to optical heating as a function of incident laser polarization and other parameters: (a) substrate temperatures, $T$, including a fit (black line) to the form $C_1 \cos^2 q + C_2$; (b) oxide thicknesses, $t_{oxide}$, (c-d) nanowire widths, $w$, and materials: (c) Au and (d) Au/Pd. Angles are defined as in Figure 1a.

The polarization dependence, with larger response in the transverse polarization, is consistent with a significant contribution from resonant plasmonic heating, rather than direct absorption by the metal. Previous work has demonstrated the existence of a strong, dipole-active transverse plasmon mode for nanowires and nanobelts;[35-37] calculations for our geometry confirm this transverse dipolar tuning response (Figure S2). When exciting this mode, one would expect comparatively enhanced heating for transverse polarized light relative to polarizations not aligned with the plasmon resonance. The data in Figure 2a-b is very well described with a $-\Delta G = C_1 \cos^2 \theta + C_2$ dependence, as shown with the black line in Figure 2a. This is consistent with previous work on plasmonic nanorods,[38] confirming the dominant plasmonic contribution.

The difference in the magnitude of the conductance change between the samples on differing oxide thicknesses and substrate temperatures indicates that thermal conduction in the oxide layer limits the thermal path and sets the steady-state temperature rise. With a thinner oxide layer, the thermal resistance decreases and reduces the heat-induced change in conduction (Figure 2b). At higher temperatures the thermal resistance of $SiO_2$ decreases[39] (as does that of the metal electrodes), resulting in a smaller heat-induced change in conductance, $-\Delta G$ (Figure 2a). Through the measured temperature dependence of the wire resistance, we quantify the photoinduced temperature increases (see Fig. 4 below).

Polarization dependent measurements on a series of wires with different widths confirm the resonant plasmonic contribution to the heating. The polarization dependences for the two different wire widths are quite different (Figure 2c), with the 130 nm wide nanowire having a $-\Delta G$ with a strong polarization dependence, while the $-\Delta G$ of the narrower 70 nm nanowire is nearly isotropic. Direct, non-plasmonic absorption in metal is polarization-independent; since the transverse plasmon resonance of the 70 nm width structure is non-resonant at 785 nm, this device shows only slight polarization dependence.

For nanowires with extremely damped plasmons, fabricated with an Au/Pd (40% Pd) alloy rather than pure Au, there is no observable polarization dependence to the heating response. The Au/Pd blend was selected to allow the fabrication of nanowire devices with comparable physical geometries, but with additional damping in the metallic dielectric function. Alloy wires with two different widths, 165 nm and 95 nm, were examined and the dependence of $-\Delta G$ on incident polarization characterized (Figure 2d). The polarization dependence of $-\Delta G$ for the 165 nm

Au/Pd nanowire is similar to the 70 nm Au nanowire, varying only in amplitude. This contrasts strongly with the analogous Au devices, which showed a significant increase in $-\Delta G$ for the wider nanowire when exciting the resonant, transverse mode. The nearly polarization-independent value of $-\Delta G$ for both alloy nanowires illustrates that the transverse plasmon resonance has been eliminated, and only direct absorption by the alloy determines the overall thermal response.

The photothermal response of the nanowires was modeled using the finite element method (FEM, COMSOL 4.3b). The geometry of the simulated structure closely matched the measured dimensions of fabricated devices. The nanowire section measured 500 nm in length, with a pad divergence angle of 45 degrees, and a radius of curvature of 150 nm at the wire/pad junctions. The 15 nm Au layer forming the rod and pads was modeled using an experimentally derived dielectric function[40] for evaporated gold. The FEM model was solved in three steps: (1) Ohmic absorption in the Au was calculated using Maxwell's equations, (2) the steady-state temperature was determined by solving the heat equation for conductive transport, and then (3) the current density within the wire assuming a DC bias. The optical model employed an effective medium approximation with neff = 1.25 to account for substrate effects, and perfectly matched layers (PMLs) to absorb scattered light at the boundaries. The background field was a Gaussian beam focused to 1800 nm FWHM, with a power of 2.2 mW, to match the experimental excitation parameters. The second calculation step determined the thermal response by using the optically-induced Ohmic heating within the nanowire to drive a steady-state temperature gradient. For this thermal model the modeled geometry included all elements of the experimental system, with the Au layer supported by a 200 nm SiO2 layer on silicon. The thermal conductivities used to model the three constituent materials were: $k_{Au}$=318, $k_{silica}$= 1.38, and $k_{Si}$=163 J/(kg-K). The outer boundaries of the simulation space were held constant at $\Delta T = 0$ K. The last simulation step determined the steady-state current densities within the wire for DC conductance. Weighting the local temperatures by the local current densities, $\mathbf{J}$, permits a direct comparison between the calculated temperatures and the experimental values:

$$\Delta T = \frac{\iiint \Delta T(x,y,z)|J(x,y,z)|dV}{\iiint |J(x,y,z)|dV} \tag{1}$$

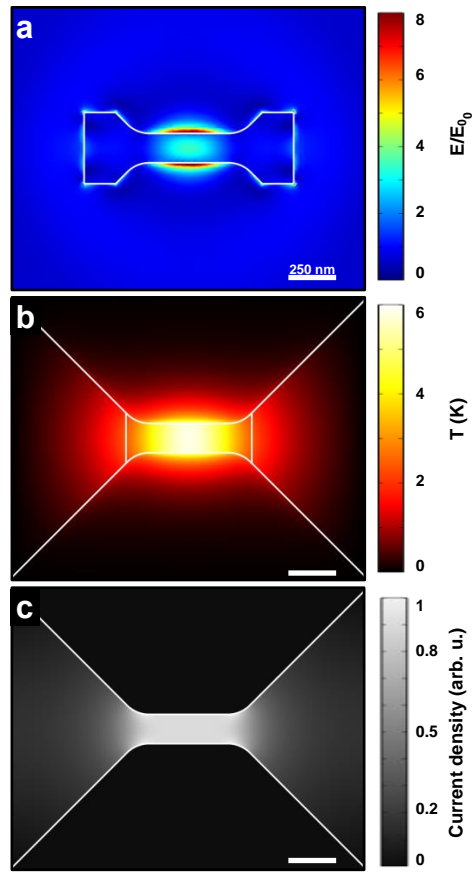

**Figure 3.** Finite element method (FEM) simulation steps for calculating plasmonic heating in a nanowire. (a) The electromagnetic response is calculated for a finite structure embedded in an n=1.25 effective medium, under Gaussian illumination (2.2 mW, 1800 nm FWHM). (b) The induced temperature change resulting from absorption within the central wire is calculated

assuming steady-state thermal transport. (c) The current density is calculated, as a weighting function for the local temperatures, to allow comparison with the experimental results. All scale bars are 250 nm.

The photoresponse allows the quantitative determination of the average temperature rise due to optical heating. Temperature dependent resistance, $R$, measurements were performed on 14 single nanowire devices. The typical dependency of $R$ as a function of temperature, $T$, for a standard Au nanowire on 0.2 μm of Au is plotted in Figure 4a. All devices showed similar linear behavior with slopes, $dR/dT$, ranging from $0.09 - 0.13$ Ω/K. These values were used to calculate the plasmonic heating.

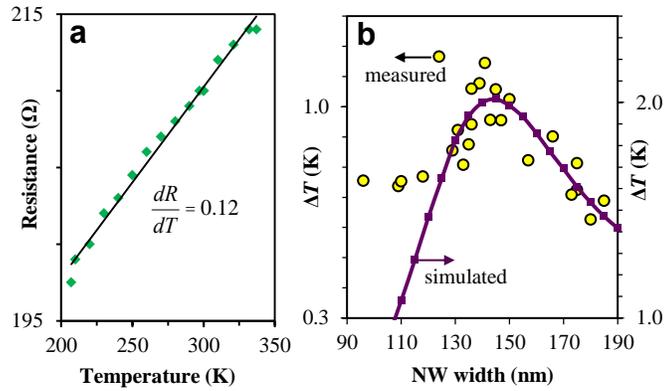

**Figure 4.** (a) Typical resistance as a function of temperature dependence for a standard gold nanowire. (b) Calculated and simulated plasmonic heating as a function of nanowire width for sample at $T = 300$K on a substrate with 0.2 μm thick oxide.

For a fixed bias voltage, Ohm's law relates a small change in the current to a small change in the resistance $dR = -\frac{R^2}{V}dI$, and the change in temperature can be inferred from $dT = dR/\left(\frac{dR}{dT}\right)$. With these equations, the change in temperature, or heating, can be quantified through the expression:

$$\Delta T = \frac{-R^2 dI}{\frac{dR}{dT}V_{dc}} = \frac{-R^2 \Delta G}{\frac{dR}{dT}} \quad (2)$$

Using this equation, the $\Delta T$ due to plasmonic heating in the transverse direction was determined to range from 0.6 – 1.2 K with 0.2 µm substrates at 300 K. Decreasing the substrate temperature to 80 K, which increased the thermal resistance between the device and the substrate, typically doubled the temperature change, making $\Delta T \sim 2$ K. A thicker oxide (2 µm) thermally isolates the device to a greater extent, and temperature changes up to 4.5 K have been measured for such substrates held at 80 K.

Plasmonic heating as a function of nanowire width shows a strong dependence on the plasmon resonance (Figure 4b). The theoretical and experimental width dependences agree well, with the calculated temperature changes (right axis) slightly larger than the measured values (left axis). This discrepancy could be the result of experimental factors not present in the simulations, such as Au plasmon damping by the Ti adhesion layer[41] or surface roughness. Discrepancies between the experimental and theoretical values most likely arise from the use of bulk material constants in the calculations. These constants are known to be altered for materials where the physical dimensions approach the underlying length scales determining properties such as electrical conduction and thermal diffusion.[42,43] Interestingly, the maximum temperature rise for a nanowire does not correspond precisely with the maximum plasmon absorption amplitude at the laser line. Instead, the maximum $\Delta T$ is calculated predicted for a slightly narrower wire, due to such a geometry's with a larger thermal resistance (Figure S3).

Plasmonic heating in single Au nanowires has been quantified using through electronic transport measurements under illumination. This method has allowed us to quantify the temperature changes in individual nanowires, and observe the relative contribution of plasmon heating to the total temperature increase for both resonant and nonresonant devices. Nanowires with widths ranging from 100 – 180 nm have shown temperature increases up to 4.5 K with a light intensity of 20 kW/cm$^2$. Polarization dependent measurements confirm the role of plasmons in the photoheating process, with light polarized to drive a plasmon mode producing a temperature rise as much as 2.5 times higher that associated with nonresonant absorption.

These experiments imply that the steady-state increases in electrode temperature due to laser illumination in previous experiments[33–35] could have been at most ~2 K, with the dominant outflow of heat taking place via electronic conduction in the nanowire, eventually followed by phonon-based thermal transport through the substrate. This emphasizes the importance of large-area contacts when designing structures that minimize local photothermal effects. Conversely, maximizing photothermal effects for supported nanostructures requires minimal electrical connectivity to the macroscopic world (to avoid electronic heat flow), high thermal resistance substrates (amorphous oxides or suspended structures), and high quality plasmon resonances (avoiding damping due to defects in metal structure or purity). Moreover, this work has demonstrated a working nanoscale, polarization dependent, plasmonic bolometric device. Such devices could be useful for future nano-optical applications and technologies.

AUTHOR INFORMATION

**Corresponding Author**

*E-mail: natelson@rice.edu

**Author Contributions**

#These authors contributed equally to this work.

**Notes**

The authors declare no competing financial interest.

ACKNOWLEDGMENT

The authors would like to thank Prof. Naomi Halas for helpful considerations. JBH would like to acknowledge the assistance of Yajing Li and Kenneth Evans and the financial support from the Robert A. Welch Foundation postdoctoral fellowship and the Lockheed Martin Corporation through LANCER. DN acknowledges Robert A. Welch Foundation grant C-1636. MWK acknowledges the support of the Robert A. Welch Foundation (grant C-1220) and the DoD NSSEFF (N00244-09-1-0067).

TOC Figure:

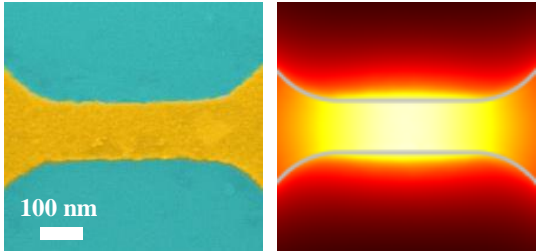

# Thermoplasmonics: Quantifying Plasmonic Heating in Single Nanowires

*Joseph B. Herzog[1,2,#], Mark W. Knight[3,4,#], Douglas Natelson[1,3,a)]*

[1]Department of Physics and Astronomy, Rice University, Houston, Texas, 77005, United States

[2]Department of Physics, University of Arkansas, Fayetteville, Arkansas, 72701, United States

[3]Department of Electrical and Computer Engineering, Rice University, Houston, Texas, 77005, United States

[4]Laboratory for Nanophotonics, Rice University, Houston, Texas, 77005, United States

\* Corresponding Author: Douglas Natelson:

  Email: natelson@rice.edu

# Authors contributed equally

# Supplemental Figures

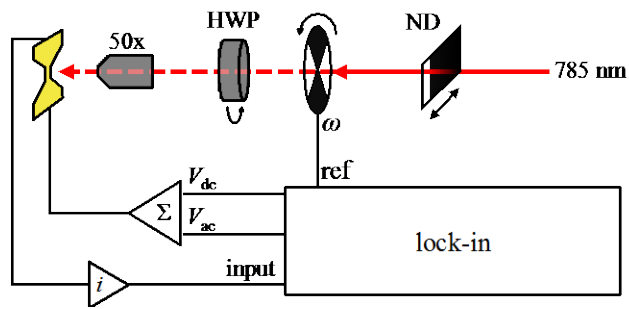

**Figure S1.** Schematic diagram of the experimental setup. A 785 nm continuous wave laser that can be scanned across sample is incident on single nanowire connected to triangular electrodes. The laser is focused using a 50x objective with NA=0.70, with its intensity and polarization controlled by a neutral density filter and a half-wave-plate, respectively. The laser is modulated with a chopper to allow lock-in detection at the modulation frequency. A constant voltage (dc) and a reference voltage (ac) are summed and applied across nanowire; the resulting current is amplified and then measured with the lock-in amplifier.

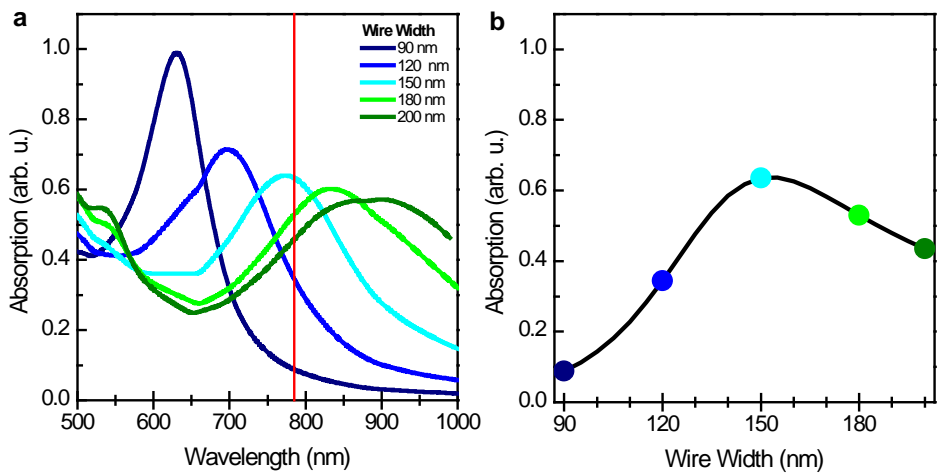

**Figure S2.** **(a)** Calculated absorption spectra of Au nanowires, with the 785 nm laser line indicated by the vertical red line. (b) Calculated absorption amplitudes at the laser line (black line), with the colored points corresponding to the spectra shown in panel (a). All calculations were performed in 2D for infinite metallic wires with a constant thickness of 15 nm, and widths varied between 90 nm and 200 nm. The wires were embedded in an effective medium with an index n=1.25.

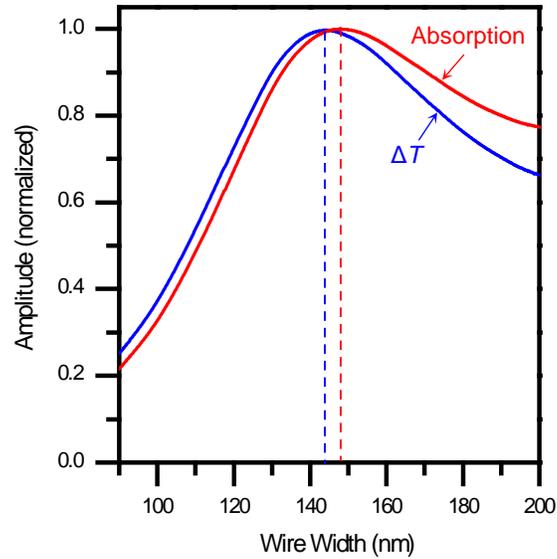

**Figure S3.** Calculated absorption in the 3D nanowire (red line), plotted vs. the calculated final averaged temperature rise (blue line). The slight shift between these two curves results from the interplay between the width dependence of the nanowire absorption, and the width dependence of the thermal resistance. Due to the increased thermal resistance of narrow wires, the maximum temperature increase occurs for slightly narrower widths than the maximum absorption. For our materials and dimensions this is a small effect, with the widths that maximize absorption and temperature differing by only ~ 5 nm.